# Natural Transfer of Viable Microbes in Space from Planets in the Extra-Solar Systems to a Planet in our Solar System and Vice-Versa

Short title: Transfer of Microbes between solar systems

**Mauri Valtonen,[1] Pasi Nurmi,[1] Jia-Qing Zheng,[1] Francis A. Cucinotta,[2] John W. Wilson,[3] Gerda Horneck,[4] Lennart Lindegren,[5] Jay Melosh,[6] Hans Rickman,[7] Curt Mileikowsky[8]***

[1] Tuorla Observatory, Turku University, FIN-21500 Piikkiö, Finland.

[2] NASA Johnson Space Center, 2101 NASA Road 1, Houston, TX 77058, USA

[3] NASA Langley Research Center, Mail Stop 111, Hampton, VA 23681-2199, USA

[4] Deutsches Zentrum für Luft- und Raumfahrt, Linder Höhe, D-51147 Köln, Germany

[5] Lund Observatory, Lund University, Box 43, SE-22100 Lund, Sweden

[6] Department of Planetary Sciences, Lunar and Planetary Laboratory, University of Arizona, 1629 East University Boulevard, Tucson, AZ 85721-0092, USA

[7] Department of Astronomy and Space Physics, Uppsala University, Box 515, SE-75120 Uppsala, Sweden

[8] Royal Institute of Technology (KTH), SE-10044 Stockholm, Sweden

* Curt Mileikowsky deceased 24.6.2005. He directed this research until his death.

**To whom correspondence should be addressed:**

Dr. Mauri Valtonen

FIN – 21500 Piikkiö

fax : +358 2 3338215

e-mail : mvaltonen2001@yahoo.com

ABSTRACT

We investigate whether it is possible that viable microbes could have been transported to Earth from the planets in extra-solar systems by means of natural vehicles such as ejecta expelled by comet or asteroid impacts on such planets. The probabilities of close encounters with other solar systems are taken into account as well as the limitations of bacterial survival times inside ejecta in space, caused by radiation and DNA decay. The conclusion is that no potentially DNA/RNA life-carrying ejecta from another solar system in the general Galactic star field landed on Earth before life already existed on Earth, not even if microbial survival time in space is as long as tens of millions of years.  However, if the Sun formed initially as a part of a star cluster, as is commonly assumed, we cannot rule out the possibility of transfer of life from one of the sister systems to us. Likewise, there is a possibility that some extra-solar planets carry life that originated in our solar system. It will be of great interest to identify the members of the Sun's birth cluster of stars and study them for evidence for planets and life on the planets. The former step may be accomplished by the GAIA mission, the latter step by the SIM and DARWIN missions. Therefore it may not be too long until we have experimental knowledge on the question whether the natural transfer of life from one solar system to another has actually taken place.

*Subject headings*: astrobiology --- stellar dynamics --- meteors, meteoroids --- (stars:) planetary systems

1. INTRODUCTION

All presently known life on Earth is controlled by DNA/RNA for information and replication, using the same code in a complicated manner, which gives a strong indication that all life on Earth has a common ancestor cell. This generally is the opinion of biologists today, and it is usually assumed that the ancestral cell originated on Earth itself. Sometimes the question is raised, however, whether that cell, or one of its forerunners, had arrived from elsewhere, by some natural way of transfer from space. It has been earlier shown (Gladman 1997, Mileikowsky et al. 2000) that the transfer to the Earth of viable, dormant bacteria inside ejecta expelled into interplanetary space by comet or asteroid impacts on Mars would have been possible, particularly during the first 0.7 Gyr of our planetary system, and also, to a lesser extent, during the following 4 Gyr. This indicates that if DNA/RNA life had once originated on early Mars – planet Mars cooled before the Earth and probably became habitable before the Earth – it could then have been transferred to the Earth. It is also possible that bacterial life on Earth would have found a temporary refuge in space on board of meteoroids while impact-induced extinction was taking place at Earth's surface (Sleep and Zahnle 1998, Wells et al. 2003, Gladman et al. 2005). Here we investigate whether it is possible that our ancestral cell might have come from a planet or a large moon in an extrasolar system, surviving the severe conditions during the journey and travel time of tens of millions of years.

2. EJECTA FROM PLANETS

When a comet or an asteroid impacts a planet, a crater is formed, the diameter of which is a known function of the diameter of the impactor and of its velocity. Part of the impactor and of the ground it hits is heated by a shock wave to melting point and it evaporates. Most of the ejecta resulting from the impact are heated to temperatures greater than 100°C, lethal for bacteria inside; however, a few percent have been found to be weakly shocked or unshocked, with temperatures < 100°C (Melosh 1989, Artemieva and Ivanov 2004, Fritz et al. 2005). They come from a shallow layer just below the planet's surface within a few projectile diameters of the point of impact, well within the final crater rim, called the spall layer, where shock waves almost cancel each other out. A number of ejecta from the spall layer achieve speeds higher than escape velocity from the planet – 5 km/sec for Mars and 11 km/sec for Earth – and thus leave the planet to orbit in interplanetary space. During their way up through an atmosphere of a pressure of one or a few bar, the outermost skins of the ejecta melt and then solidify after leaving the atmosphere. The passage up through the atmosphere is so rapid that the heat does not have time to spread into the ejecta's inner parts, which remain at temperatures < 100°C for ejecta larger than a certain size, ~ 20 cm (Clark 2001, Horneck et al. 2001, Burchell et al. 2001, 2003, 2004, Willis et al. 2006, Fureby, C., Almström, H., private communication.). A fraction of these larger ejecta are expelled by the gravitational influence of Jupiter-sized planets into interstellar space. They are the vehicles with the potential to transport viable bacterial spores in interstellar space for times as long as they can survive the threats there.

What is the approximate number of ejecta of different sizes leaving Earth and Mars, respectively, during the early 0.7 Gyr and the last 4 Gyr ? The number of ejecta expelled from Mars or Earth into all directions by impactors of diameters 1-20 km or higher can be calculated on the basis of solar system evidence. Markings were left on the planets and moons in our Solar System by impacting comets and asteroids in the form of impact craters of various diameters, revealing the sizes and the size distribution of the impactors. But weathering on the Earth and Mars has hidden or obliterated many early impact craters, and thus a direct count of detectable craters would underestimate the number of actual impacts. On the Moon however, with its lack of atmosphere, water, wind and erosion, all imprints remain unaltered to this day, and lunar crater counts include even the earliest impacts of the heavy bombardment. By correcting for differences in gravity and size, the effects on Mars and Earth have been deduced and the impact history of both planets calculated: from the frequency distribution of impact crater diameters and shapes, it has been possible to calculate the frequency distribution of the impactor diameters and then, for each impactor size and speed, the frequency distribution of ejecta size. The ejecta size is important because of their ability to achieve radiation shielding against the Galactic Cosmic Ray nuclei. It requires ejecta with radii larger than 0.2 m, and the survival increases with ejecta size. The surfaces of ejecta get heated, melt and partly ablate while moving up at escape velocity, or more, through an atmosphere of ~ 1 Bar. The larger the ejecta, the larger the portion that remains < 100°C.

Approximately $4 \cdot 10^{12}$ ejecta of this size range with T < 100°C were ejected on average, in all directions, from Mars' surface during the last 4 Gyr by impactors of diameters 1 - 20 km, i.e. an average ~ $10^9$ per Myr (Mileikowsky et al. 2000). The rate of ejection *R* (per Myr) of bodies whose size (radius) is greater than $l$ is a decreasing function of the ejecta size (see column VI of Table 1):

$$R(l) \approx 3 \times 10^7 l^{-4}, \qquad (1a)$$

when $l > 0.7$ m. The distribution is flatter ($R \sim l^{-1}$) for smaller bodies. Of these, ~ 3 – 10 % made their way to the interstellar space (Gladman et al.2000). During the first 0.7 Gyr the bombardment of the planets was much heavier, perhaps a thousand times heavier (Chyba et al. 1994). For our discussion here it is not essential whether the bombardment was steady or happened in a spike at the end of the period, as suggested by recent simulations (Gomes et al. 2005). Therefore during the period of heavy bombardment the rate of sending out ejecta to the interstellar space is:

$$R(l) \approx 2 \times 10^9 l^{-4}. \qquad (1b)$$

Only those ejecta which have not spent their maximal life-carrying time inside the planetary system before being expelled into interstellar space can serve as potential life carriers to another solar system. Approximately 50% of all ejecta leave the Solar system within 50 Myr, the rest take a longer time (Melosh 2003). This reduces the number of potentially life-carrying bodies in the interstellar space given in Eqs. 1a and 1b somewhat.

3. SURVIVAL OF BACTERIA IN SPACE

The main threats to bacterial survival inside ejecta are effects by short time (ns) exposure to high temperatures up to several 100 ºC and shock pressures up to 50 GPa during launch and exit through atmosphere, and during arrival and landing on another planet; when travelling in space, radiation in the form of Galactic Cosmic Ray and Solar Cosmic Ray particles and natural radioactivity from the ejecta material itself; the DNA decay, e.g. caused by hydrolysis leading to release of DNA bases, and exposure to excess temperature over sustained time or extended periods in vacuum.

All these effects have been discussed in depth in Mileikowsky et al. (2000). The survival times against the different radiation threats refer to existing bacterial species with the highest radiation resistance known, *Deinococcus radiodurans* as well as *hyperthermophile Archaea*. Bacterial endospores are also interesting candidates. *D. radiodurans* efficiently coordinates the recovery from ionizing radiation through a complex network of a highly condensed nucleoid structure, DNA repair and metabolic pathways (Cox and Battista 2005). As to the occurrence of DNA damage induced by the different threats, *D. radiodurans* is not significantly superior to many other bacterial species. In *B. subtilis* spores the DNA is extremely well protected against environmental stressors, such as desiccation and radiation, mainly because of three factors: (i) a dehydrated cytoplasma enclosed in a thick protective envelope (Nicholson et al. 2000), (ii) the stabilisation of their DNA by specific small proteins, and (iii) efficient DNA repair pathways during germination (Moeller et al. 2007). In shock recovery experiments, the spores survived simulated impact scenarios at shock pressures up to 40 GPa, which corresponds to the pressures determined for Martian meteorites (Horneck et al. 2008). These two microorganisms were taken as model systems for our calculations, because they indicate the upper limit of microbial resistance against radiation.

If DNA decay by hydrolysis or vacuum were effective, the viable transfer from planets in extrasolar systems to the Earth would be impossible. The survival time of bacteria due to DNA damage by hydrolysis is of the order of hundred thousand years, not long enough for long space flights considered here, based on available experimental evidence (Nakamura et al. 1998, Lawley and Brookes 1968, Karran et al. 1980). In *vacuated* ejecta volumes the DNA decay by vacuum-caused damage occurs much faster than damage by hydrolysis or radiation, in tens of years (Dose et al. 1991, Horneck 1993, Horneck et al. 1994). For protection against vacuum, the melting of the outer layers of the ejecta during the passage up through the source planet's atmosphere is important. It should create a tight shell which solidifies on leaving the atmosphere. As long as there are no cracks going all the way through the shell, the bacteria containing pores inside are tightly isolated from the vacuum outside and the atmosphere does not leak out.

The Galactic Cosmic Ray survival calculations were made using the NASA HZETRN code (Cucinotta et al. 1995, Wilson et al. 1995) supported by experimental checking of bacterial survival at the large heavy particle accelerator at the GSI, Darmstadt, Germany (Baltschukat and Horneck 1991). The code covers all Cosmic Ray nuclei up to nickel, with $Z = 28$. Each element nuclei have their own energy spectrum and intensities in interplanetary space, with even higher intensities of the soft – but damaging – components in interstellar space. They are used in the calculations for energies up to 100 GeV/amu. The calculations regarding natural radioactivity in the ejecta material used the same code for the alphas from the known concentrations of uranium and thorium, and for the activity from potassium at the known concentrations in the Martian meteorites and in Earth materials.

Both the number of projectiles $R$ sent to the interstellar medium and the survival time $t$ (in Myr) of bacteria inside them depend on the size of the projectile $l$ (in m). The former dependence was given above (Eqs 1a, 1b). Discounting the exposure to vacuum and hydrolysis, the survival time is (see column IV of Table 1):

$$t \approx 75 l^2 . \qquad (2)$$

Thus, the product $Rt^2$ is rather constant since the lifetime of bacteria inside larger bodies increases with body size, while at the same time the number of ejected bodies decreases as body size increases (Eq. 1b). We may write

$$Rt^2 \approx 10^{13} \qquad (3)$$

which is valid within a factor of two for the range of bodies between 0.7 m and 7 m, the relevant range here.

4. NUMBER OF LIFE-CARRYING TRANSFERS

In our calculations it is convenient to use 1 Myr ( $= 3.156 \times 10^{13}$ s) as the time unit and 1 pc ( $= 3.086 \times 10^{13}$ km) as the distance unit; the velocity unit is then 1 pc/Myr = 0.978 km/s $\approx$ 1 km/s. The expected rate of transfers of life-carrying bodies from other solar systems to the Earth depends on conditions such as the mean density of ejecting systems in the solar neighbourhood and their velocity distribution. Although the present characteristics of the solar neighbourhood are fairly well known, previous conditions, and especially those of the early history of the solar system, are essentially unknown. However, it is likely that our Sun, and the Solar System, was formed together with many other stars and their planetary systems from a common molecular cloud, and that these systems existed as star clusters or association during a considerable length of time, on the timescale of a few hundred million years. Therefore, two alternative conditions were considered: the Sun was a member of a star cluster similar to the clusters that are observed at the present time, and the current solar neighbourhood conditions.

We assume that the stars have an isotropic velocity dispersion with a Gaussian probability density in each coordinate, and $\sigma$ is the standard deviation per coordinate, so that the speed $V$ of the stars relative to the Sun has a Maxwellian probability density:

$$f(V) = \sqrt{\frac{2}{\pi}}\, \sigma^{-3} V^2 \exp\left(-V^2/2\sigma^2\right). \qquad (4)$$

The ejection process is characterized not only by the rate of expulsions $R$ from the planetary systems but also by the resulting distribution of the escape velocity $v_{esc}$, i.e. the velocity at infinity of the ejecta relative to the expelling star. The escape velocity is usually small (< 1 km/s), a fraction of the orbital speed of the planet responsible for the ejection, but may reach 10 km/s in rare cases (with a probability of $\approx 10^{-4}$, Valtonen and Innanen 1982, Zheng and Valtonen 1999). The distribution of escape directions can be taken to be isotropic. Although some preference for the invariable plane is expected for a given system, this is eliminated when considering an assembly of stars with random orientations of their ecliptic planes.

In order to calculate the impact rate we need to know the probability density of $v$, the approach speed of the ejected body relative to our solar system when it is still at a large distance from us. Since $V \gg v_{esc}$, we commit only a minor error if we replace $V$ by $v$ in Eq. 4. We need only to multiply the final impact rate by a factor of $1.5 - 2$ to correct for the neglect of $v_{esc}$.

The probability that a body approaching the solar system will directly hit the Earth is exceedingly small. A hit is much more likely if the body is first captured into a bound orbit by one of the giant planets, and collides with the Earth only after many orbital revolutions. First we calculate the probability that an interstellar body comes into the sphere with a 30-AU radius. This cross-section (in $pc^2$) is

$$\Sigma_T \approx 4 \times 10^{-6} v^{-2} \qquad (5)$$

where $v$ is the velocity at infinity of the body relative to the Sun in km/s. It is assumed that the velocity $v$ is much less than 5 km/s, approximately the orbital speed of Neptune. For larger values of $v$ the cross-section is somewhat greater than what is given by Eq. 5 (Binney and Tremaine 1987). If the orbit comes inside this sphere, the interstellar body may cross the orbit of one of the giant planets and thus it may be (but not necessarily) captured. Secondly, we calculate the collision probability within this sphere. In the probability calculations, we have used orbit calculations with a second order integrator (Bulirsch and Stoer 1966) and a solar system model consisting of the Sun and the giant planets. The collision probability is obtained by applying the Öpik's equations (Öpik 1951) during the orbital evolution. The calculated collision probability with the Earth within time $\tau$ (in units of Myr) is

$$P(\tau) \approx 1.5 \times 10^{-9} \tau \qquad (6)$$

for $v \leq 0.5$ km/s and for $\tau$ up to several tens of Myr. (Note that $v$ is the velocity at a large distance from the solar system; the actual collision velocity is at least an order of magnitude greater.) The product of the expressions in Eqs. 5 and 6 gives then the effective cross-section $\Sigma$ for impact on the Earth:

$$\Sigma(v, \tau) \approx 6 \times 10^{-15} \tau v^{-2} \quad (v \leq 0.5 \text{ km/s}) . \qquad (7a)$$

For higher values of $v$ we get directly from simulations, without recourse to the intermediate steps of Eqs. 5 and 6:

$$\Sigma(v, \tau) \approx 1.5 \times 10^{-15} \tau v^{-4} \quad (v \geq 0.5 \text{ km/s}) . \qquad (7b)$$

Thus the cross-sections of Eqs. 7a and 7b refer to both the passage within 30 AU from the Sun, a capture into a heliocentric orbit and a subsequent collision with the Earth.

Let us now consider the transfer of bodies from other solar systems to our solar system with a given maximun flight time $t$. The free flight time in interstellar space is $t - \tau$ where $\tau$ is taken to include the time spent in both planetary systems, the donor and the receiver system. If $N$ is the number density of donor systems, and $R$ is the ejection rate per system, then the mean density of ejecta not older than $t - \tau$ is $NR(t - \tau)$. In the time interval $D_t$ the receiver system moves through a (collision) volume $v D_t d\Sigma$ where $d\Sigma$ is the cross-section for collision per $d\tau$ interval. This multiplied by the (viable) ejecta density and integrated over $\tau$ as well as the speed distribution $f(v)$ gives us the number of impacts on the receiver planet (e.g. the Earth):

$$n = NRD_t \int_0^\infty \left[ \int_0^t \frac{\partial \Sigma(v,\tau)}{\partial \tau}(t-\tau)\,d\tau \right] f(v)\,v\,dv \ . \tag{8}$$

According to Equations (7a) and (7b) the integral inside the brackets is $3\times10^{-15} t^2 v^{-2}$ for $v \leq v_0 = 0.5$ km/s and $0.75\times10^{-15} t^2 v^{-4}$ for larger values of $v$. The outer integral must therefore be written as a sum of two integrals, the first from 0 to $v_0$, the second from $v_0$ to infinity. Numerical integration shows that the second integral makes a minor contribution to the total for $\sigma \leq 27$ km/s. In all practical situations we therefore make only a minor error by neglecting the second integral, i.e. by neglecting impacts with $v > v_0 = 0.5$ km/s. With this approximation the final result is

$$n \approx 0.4\times10^{-15} \times (Rt^2) \times (D_t N / \sigma^2) \ . \tag{9}$$

As we mentioned above, the product $Rt^2 \approx 10^{13}$.

5. THE BIRTH CLUSTER OF THE SUN

In order to estimate the other product $D_t N/\sigma^2$ we first consider the case when the Sun was young, and was probably a member of a star cluster or association. The cluster had a much higher density of stars than in the general Galactic field, and its velocity dispersion was much lower than we observe around us now. It is in fact believed that most of the Galactic field stars originated in clusters or associations, which subsequently disrupted and dissipated into the general star field on time scales of 10 to 1000 Myr. As to star density and velocity dispersion in the proposed cluster, we may refer to the central parts of two real star clusters, assuming that the early solar system would have been located in an environment similar to them. The Trapezium cluster in the Orion nebula is one of the densest known clusters with a number density of $N_* = 2200$ stars per $pc^3$ and a velocity dispersion of $\sigma = 2$ km/s (Herbig and Terndrup 1986, Prosser et al. 1994). Even though the age of the Trapezium cluster is only about 1 Myr, with planetary systems still in the process of formation, it is expected to last about 250 Myr (Boutloukos and Lamers 2003). An older cluster, exemplified by the Hyades (estimated age 625 Myr), has a density of $N_* = 2$ solar masses per $pc^3$ and a velocity dispersion of $\sigma = 0.25$ km/s (Perryman et al. 1998). The value of $D_t$ is either the cluster lifetime or the maximum "viable" lifetime, whichever is smaller. In Table 1 the ejecta are divided in size groups 2 – 7 which contribute about equally to the "viable" ejecta. The medium lifetime for these groups is 200 Myr. It is shorter than the cluster lifetime in both of our cluster examples. Thus we may use $D_t = 200$ Myr in our calculation. The geometric mean for the product $D_t N/\sigma^2 \approx 2.5 \cdot 10^4$, within a factor of four the same for both cases.

Gravitationally bound structures similar to the Hyades cluster may be able to exist for more than 1000 Myr before they are disrupted by the tidal field of the Galaxy and passing giant molecular clouds (Wielen 1988). Remnants of disrupted clusters should be ubiquitous (De la Fuente Marcos 1998), and some of them could still be revealed by local density enhancements in the phase space of stellar motions, i.e. by the nearly parallel motions of a group of stars through space. The low velocity dispersion within such a remnant (few km/s) would increase the probability of transfer between its members. It is at present not known if the Sun belongs to a recognisable cluster remnant. If it does, the low velocity dispersion would probably be more than compensated by an even lower mean density of the remnant stars. Thus it is not likely that such structures could significantly enhance the impact probability for the current solar-neighbourhood conditions. In the general Galactic star field, the product $D_t N/\sigma^2$ is only of the order of unity, and we may neglect the possibility of current arrivals of life-carrying bodies to the Earth, in comparison with the events in the solar system birth cluster.

## 6. DISCUSSION

If we now introduce our best numbers into Eq. 9, the result is that the Earth may have suffered about 100 collision with life-carrying bodies originated from other stars. However, there are factors that could reduce or increase the number $n$ considerably. As to the possible reduction, we have to remember that in Eq. 9 $N$ should be the number density of stars in the cluster that act as a source of life carrying bodies which may be different from total number density of stars $N_*$. Generally $N$ is smaller than the actual number density of stars $N_*$. How much smaller it is depends, among others, on the fraction of Earth-like planets where life originates under favourable conditions, on the fraction of ejecta that are not carriers of "viable" life when they enter the interstellar medium, on the fraction of stars which have planetary systems, on the fraction of planetary systems which have terrestrial planets, and on the fraction of terrestrial planets orbiting in zones habitable for DNA based microbes.

In order to estimate the possible reduction factors, we may use the existing knowledge about the 250+ known extrasolar planetary systems around main-sequence stars. With the exception of about 25 multiple systems, only a single giant planet has been detected in each system. Most of these systems are within 50 pc from our Sun. The information we have today about extrasolar systems is necessarily strongly biased due to the present limitations of sensitivity of the detection methods used (Doppler effect, Transit photometry). Consequently, terrestrial-type planets are difficult to detect, and our sample is mainly composed of Jupiter- and Saturn-sized planets (and even there we are restricted to major planets with semi-major axis < 5 AU).

In case of binary stars, the components can have planetary systems if the components are far enough from each other (Innanen et al.1997). Ordinary wide binaries, with components only a few AU apart, can hardly support stable planetary orbits. Thus there is a fraction of stars which definitely are not good platforms for planetary systems with life in them. For the sake of completing the calculation, let us assume that $N \approx 10^{-2} N_*$ which may not be unreasonable.

On the other hand, $n$ could also be increased by a large factor. The rate of ejection of bodies from a planet in a hypothetic planetary system depends on the heavy bombardment in that system. So far we have taken the heavy bombardment to be the same as in our Solar System during the period 4.5 Gyr to 3.8 Gyr ago. However, the collision rate of comets on the Earth can be increased by as much as a factor of 1000 if the protective influence of Jupiter is removed (Wetherill 1994). It may well be that in most planetary systems there is no Jupiter-like protective planet and that the heavy bombardment is therefore much greater than what has been experienced in our solar system. Our calculated values of $n$ could be underestimates by a factor of hundred or so.

We may also consider different versions of the planetary systems at the receiving end for life-carrying bodies from other stars. Just to see what effect the variation of the planet arrangement may have on calculated collision rates, we have considered two additional systems : 1) an extrasolar system like ours but without Jupiter, and 2) an extrasolar system like ours but where Jupiter was assumed to be at 0.2 AU distance from the Sun. The positions and the masses of the other planets, including the third terrestrial planet, were supposed to be unchanged. Case 2 was studied because many of the planetary systems discovered so far have a Jupiter-type giant planet close to the star.

We found that the first alteration, removing Jupiter altogether, reduced the collision rate on the third planet to a level which is not significantly greater than the rate of direct collisions without the influence of major planets. Therefore we conclude that Jupiter has the effect of increasing the collision rate by a factor of about 40 above the direct (first orbital crossing) collision rate. Jupiter captures passing bodies which become long-lived members of our system and then sometimes collide with the Earth at a later time.

The second model, in which Jupiter was placed close to the Sun, had the opposite effect: the collision rate with the third planet was doubled. This is because the inner Jupiter captures passing bodies very efficiently to Earth-crossing orbits which subsequently have a high probability of collision with the Earth.

As regards capture of bodies to binary star systems, the capture is quite efficient but orbits generally do not pass close to either of the component stars. We have carried out some computer simulation experiments with a binary star system of 1 solar mass star and a 0.2 solar mass star at 400 AU separation from each other, and the binary orbit at $i = 40°$ inclination relative to the planetary system of the solar-type star. The latter was assumed to have a planetary system equal to that of our Sun. It was found that the binary component had a protective influence on the collisions on the third planet, rather like the role that Jupiter plays relative to the comet orbits (Wetherill 1994). Thus binary stars cannot help with the "reception" side of the process.

## 7. CONCLUSIONS

In summary, meteoroids above a certain "critical" size for cell viability do not become overheated throughout their interior during their passage through the atmosphere of a planet because the heat from their molten surface does not have enough time to spread to their centre before they leave the atmosphere and cool in space. With regard to heating, those meteoroids larger than "critical" size qualify as potential vehicles for viable transfer.

First consider the transfer of life in the galactic star field. Even if there are planetary systems with both a terrestrial-like planet in its habitable zone and a giant planet well positioned to expel into interstellar space ejecta from the terrestrial-like planet, the number $n$ of meteoroid ejecta of critical size or larger from all the planetary systems of the Galactic field stars which impacted Earth during Earth's first 700 million years was only $n \sim 10^{-8}$, i.e. no such meteoroid capable of carrying viable DNA / RNA based microbes impacted Earth. A different situation prevails in a cluster of young stars newly born from the same core of a collapsing molecular cloud. The short distance between the stars and their slow relative motion is favourable for an exchange of bodies and gives

$$n \approx 10^{2\pm2}. \qquad (10)$$

Even though this number is uncertain by several orders of magnitude, there is a definite possibility that bacteria carrying meteoroids of extrasolar origin have landed on the Earth. In reverse, it is possible that at least one other planetary system in our birth star cluster received a life-carrying asteroid from the Earth; and it is not excluded that the whole birth star cluster was "fertilized" in this way by live bacteria from the Earth.

What are the chances that life originated only once in the Galaxy and that by the processes described above it spread throughout the Galaxy, providing a process that is sometimes referred to as lithopanspermia (Horneck et al. 2008)?

From our discussion above, it is clear that exchanges of bacteria between planets in different solar systems are only possible during the birth cluster stage of the systems in question. As the number of life-carrying bodies received by the Earth may have been in thousands, so also other planets in other stellar systems may have received their life from other members of our original star cluster, or even from a single source, the Earth. Thus the limited form of lithopanspermia inside a star cluster is possible, while the stronger version of life spreading through the whole Galaxy from a single source could not happen via mechanisms described in this work. But life-carrying bodies originating from our solar system may have found their way to our original neighbours, and that all conditions being optimal, life seeded by our system could have spread to many other solar systems. Here in our solar system our common ancestor cell most probably originated either on the Earth or on Mars. We cannot say for sure which one since there has been millions of potentially life-carrying transfers between these two planets (Gladman et al.1996). The GAIA (Douglas et al. 2007) mission will perhaps be able to locate the members of the birth cluster of the Sun while the SIM (Unwin et al. 2008) and DARWIN (Cockell et al. 2008) missions will be able to detect planets around them and search for signs of life in the planets. Even before these missions, the currently ongoing search for life in Mars may already give an indication how likely it is that life is transported between planets by natural means.

Table 1. Maximum total survival time of *D. radiodurans* – like bacteria inside ejecta in interstellar space (ISS), and their maximum viable travel length at velocities ≤ 0.5 km/s, for seven size groups of ejecta caused by impactors bombarding a Mars-like planet, and approximate numbers of ejecta per Myr per size group.

| Size group | Radius range defining size groups (m) | Shielding column density (g cm$^{-2}$) | Maximum total survival time in ISS (Myr) $t_{ISS} = \dfrac{1.33 \cdot \ln(10^{15})}{\sigma F / Myr + 0.075 / Myr}$ | Maximum viable travel length in ISS at $v = 0.5 km/s$ (pc) | N° of ejecta (T≤100°C) from a Mars-like planet, caused by impactors in the range 0.5 – 20 km, per Myr |
|---|---|---|---|---|---|
| I | II | III | IV | V | VI |
| 1 | 0.00 – 0.03 | 0 - 10 | 12 – 15 | 6 – 7.5 | Burn in atmosphere |
| 2 | 0.03 – 0.67 | 10 – 200 | 15 – 40 | 7.5 – 20 | Many of size group 2 burn in atmosphere  9.4 × 10$^8$ |
| 3 | 0.67 – 1.00 | 200 – 300 | 40 – 70 | 20 – 35 | 3.0 × 10$^7$ |
| 4 | 1.00 – 1.67 | 300 – 500 | 70 – 200 | 35 – 100 | 1.8 × 10$^7$ |
| 5 | 1.67 – 2.00 | 500 – 600 | 200 – 300 | 100 – 150 | 2.8 × 10$^6$ |
| 6 | 2.00 – 2.33 | 600 – 700 | 300 – 400 | 150 – 200 | 1.5 × 10$^6$ |
| 7 | 2.33 – 2.67 | 700 – 800 | 400 – 500 | 200 – 250 | 1.0 × 10$^6$ |
|  |  |  |  |  | Σ = 10$^9$ |

Note : In the formula $t_{ISS}^{max} = \dfrac{1.33 \cdot \ln(10^{15})}{\sigma F / Myr + 0.075 / Myr}$, $\ln(10^{15})$ gives the dependence on the original number of bacteria in the batches per ejecta expelled into ISS ; $\dfrac{\sigma F}{Myr}$ gives the effect of the Galactic cosmic rays ; $0.075/Myr$ the effect of natural radioactivity (see Mileikowsky et al. 2000 for details).